\def\alphal{{\alpha_{\mbox{\tiny\it L\'evy}}}}

\documentclass{PoS} \title{Correlation Probes of a QCD Critical Point} \usepackage{amsmath} \ShortTitle{Correlation Probes of a QCD Critical Point}

\author{Tam\'as Cs\"org\H{o}
\thanks{Supported by OTKA grants T49466 and NK 73143, and by a HAESF Senior Leaders and Scholars Award.}\\
	Department of Physics, Harvard University, 17 Oxford St, Cambridge, MA 02138, USA\\
        MTA KFKI RMKI, H-1525 Budapest 114, POBox 49, Hungary \\
        E-mail: \email{csorgo@rmki.kfki.hu}}

\abstract{
Critical opalescence is a characteristic experimental signature of a second order phase transition in solid state physics. A new, experimentally accessible measure of opacity and of attenuation length in heavy ion reactions is suggested, as a combination of HBT radii and nuclear modification factors. This opacity is maximal when $\sqrt{s_{NN}}$, the system size and centrality correspond to the critical point of QCD. To characterize the phase transition at this critical  point, the critical exponent of the correlation function can be determined by measuring the L\'evy index of stability of the Bose-Einstein or HBT correlations. The exponent of the correlation length can be determined from fits to the multiplicity distribution in various pseudorapidity intervals, also as a function of colliding energy, system size, centrality and (chemical) freeze-out temperature. These two critical exponents determine the remaining four critical exponents and the universality class of this second order phase transition. As a control experiment, the determination of the critical exponent of the specific heat capacity is proposed, from event-by-event fluctuation measurements. To measure opacity precisely, well calibrated high transverse momentum probes are needed, such as given by the excitation function of $\gamma$ + jet correlation functions.
 }

\FullConference{High-pT Physics at LHC - Tokaj'08\\
		 March 16 - 19 2008\\
		 Tokaj, Hungary}

\begin{document}

\section{Introduction: five important milestones in Au+Au collisions at RHIC} 
The RHIC heavy ion program created a paradigm shift in the exploration of the phase
diagram of strongly interacting, hot and dense hadronic matter: instead of discussing a gas-like, weakly coupled
plasma of quarks and gluons, we now explore the properties of a perfect fluid, where  valence quarks appear as active 
degrees of freedom, and we discuss superfluidity at extremely high temperatures on the scale of $T \simeq 2 \times 10^{12}$ K.
An important first milestone in this program was a PHENIX discovery of suppression of particle production with large transverse 
momentum in central Au+Au collisions at $\sqrt{s_{NN}} = 130 $ GeV, which apparently was the first glimpse
on a new phenomenon~\cite{Adcox:2001jp}, never seen before in high energy nuclear collisions.
Soon the effect was found at higher energies too~\cite{Adler:2003qi},
and in a control d+Au experiment all the four RHIC collaborations reported the 
absence of suppression of particle production at high transverse momentum
~\cite{Back:2003ns,Adler:2003ii,Adams:2003im,Arsene:2003yk}. 
The angular correlation plot of STAR indicated particularly clearly, 
that the most energetic, near-side jets have similar angular distributions in p+p, d+Au and Au+Au collisions,
while the momentum-compensating back-to-back jets are seen only in p+p and d+Au collisions, but are 
absent  in central Au+Au collisions.
This absence of the back-to-back or away-side peak but the presence of the near-side peak 
indicates that jet suppression in Au+Au collisions is not an initial state phenomenon, but a final state effect.
These results convincingly demonstrated that a new form of matter is seen in central Au+Au
collisions, that is so opaque that it attenuates even jets of  particles with high transverse momentum,
and can be considered the second milestone of the RHIC program.
The third milestone has apparently  been  the publication of "White Papers" by all the four RHIC collaborations,
that summarized the conclusions from the first three years of data taking at RHIC,
and pointed out that the matter created in central Au+Au collisions does not behave like
a gas of weakly interacting quarks and gluons, instead it flows like a perfect fluid of 
strongly interacting quarks and gluons ~\cite{Arsene:2004fa,Back:2004je,Adams:2005dq,Adcox:2004mh}.
This consensus result later on became known as the discovery of a perfect liquid at RHIC,
and it  has been selected as the Top Physics Story of
the American Institute of Physics in 2005, the World Year of Physics~\cite{AIP:2005top}.
Among the many additional milestones to come, let me highlight two, particularly important  discoveries,
and consider the discovery of the quark number scaling of the elliptic flow as Milestone 4. 
This discovery can be interpreted in such a way, that the flowing matter has
quark degrees of freedom. 
STAR has pointed out an approximate quark number scaling of the elliptic flow in terms of the  transverse momentum $p_t$
~\cite{Adams:2005zg}, while PHENIX strengthened this argument significantly by pointing out that the quark number scaling
becomes perfect at low transverse momentum if instead of $p_t$, the transverse kinetic energy
$KE_T = m_T - m $ is used~\cite{Adare:2006ti} (where the transverse mass is introduced as $m_t = \sqrt{m^2 + p_t^2}$).
At mid-rapidity, this transverse kinetic energy is a reasonably good 
approximation of the (dimensionless and in general, rapidity dependent)
scaling variable $w$, found in first in  exact solutions of perfect fluid fireball 
hydrodynamics~\cite{Csorgo:2001xm}. 
When plotted in terms of
the universal scaling variable $w$, all presently known elliptic flow data at RHIC  
collapsed to the theoretically predicted, 
universal scaling curve~\cite{Csanad:2005gv}, 
confirming the perfect fluid picture. As the relationship between the scaling variable $w$
and the elliptic flow $v_2$ is approximately linear for $v_2 \ll 1$, this universal scaling
law is approximately invariant for a rescaling by the number of quarks.
By now, this  transverse kinetic energy scaling of $v_2$ with the number of valence quarks
has been extended to even charm and strange mesons and baryons.
The scaling of the elliptic flow with the number of valence quarks indicates experimentally, that
in this perfect fluid valence quarks are relevant, active degrees of freedom. 
Such an active role of quarks has been expected, but it has not been demonstrated experimentally 
before~\cite{Adams:2005zg,Adare:2006ti}.  This is why I consider this observation 
as the fourth important milestone at RHIC.

As a next step, it was natural to ask~\cite{Riordan:2006df}, how perfect is this perfect fluid? 
Can we measure its kinematic shear or bulk viscosity? What about its heat conductivity?
The first answer to this question was surprising to many: The measured kinematic shear viscosity
~\cite{Adare:2006nq} in Au+Au collisions at RHIC proved to be significantly below that 
of the same value for superfluid $^4$He at the $\lambda$-point~\cite{Lacey:2006bc,Zajc:2007ey}.
At the same time, the temperatures measured in Au+Au collision at RHIC were the hottest temperatures ever measured
in experiments,  of the order of $2 \times 10^{12}$ K,
in sharp contrast to the ultra-cold temperatures of superfluid $^4$He,
of the order of 4 K. In this sense, we may discuss high temperature superfluidity as the fifth important
milestone discovered in Au+Au collisions at RHIC~\cite{Csorgo:2008pe}.

These discoveries and important milestones are far from being a complete and exhaustive summary of discoveries made in
the RHIC proton-proton and heavy ion program.  
They were highlighted here in order to demonstrate, that a new form of matter is created in heavy ion 
collisions at $\sqrt{s_{NN}} $ = 200 GeV energies. This matter is so dense and so opaque, that it absorbs about 80 \% of the
jets with high energy; it  has temperatures that are higher, than any other temperature measured before; 
it exhibits high temperature superfluidity i.e. flows within errors perfectly;
even the relatively rare and heavy strange and charm quarks flow with this matter.
These properties are apparently absent at lower energies, so it is natural to ask, at what bombarding energies,
what type of collisions and at what kind of centralities do these phenomena show up first, and what are the characteristic
properties of these transitions? Can we map out the phase diagram of QCD?

\begin{figure}[t]
\begin{center}
\includegraphics[width=0.8\textwidth]{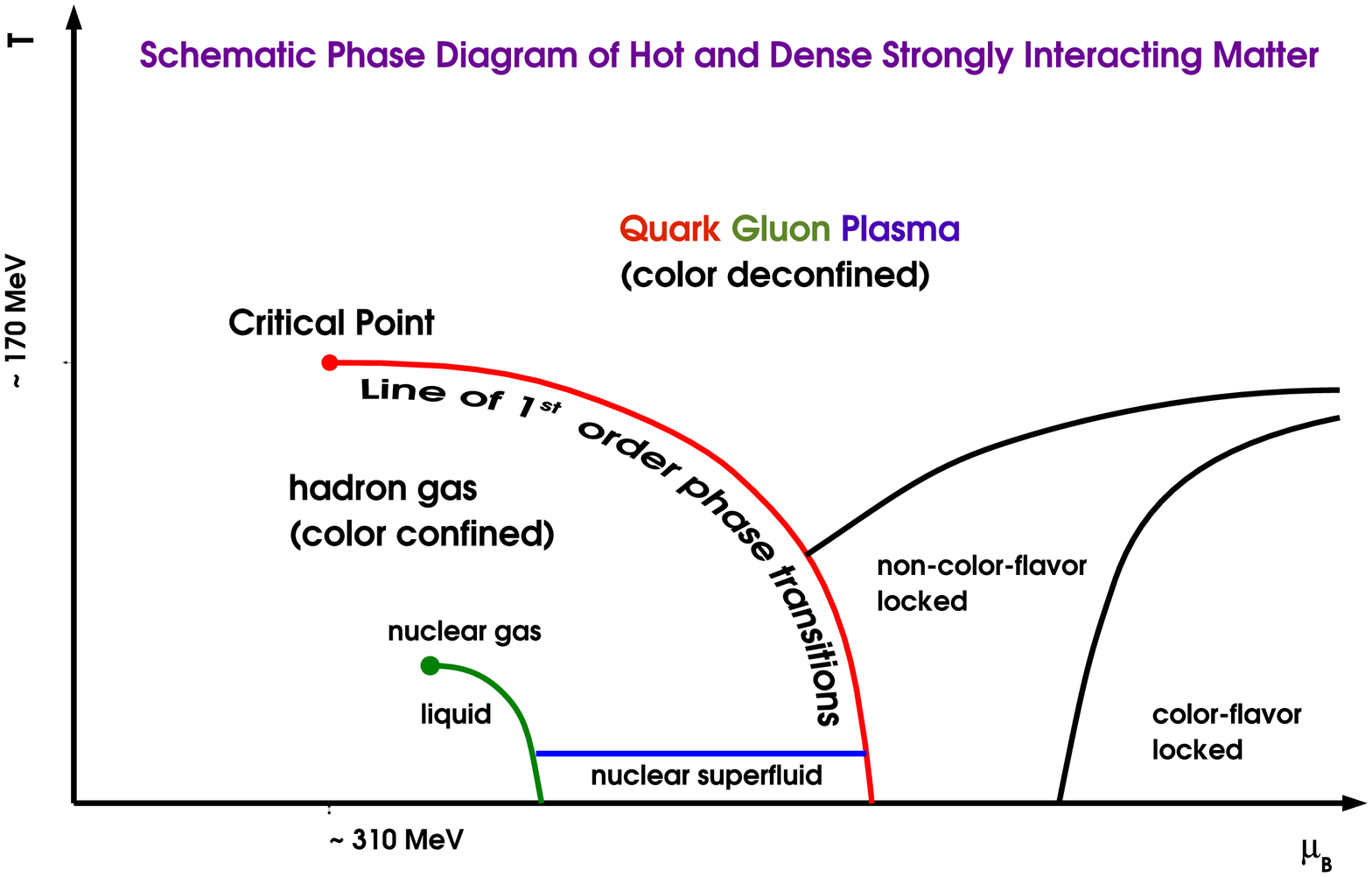}
\end{center}
\caption{%
\label{f:QCD-phase-diagram}
Schematic, theoretically expected  phase diagram of QCD.
The critical point is shown as  the end of the line of first order deconfinement
phase transitions, based on ref.~\cite{Alford:2007xm}.
Ref.~\cite{Fodor:2004nz}
predicted  $ T_E = 164 \pm 2$ MeV and $\mu_E = 360 \pm 40 $ MeV
for the location of the
end point of the line of first order QCD deconfinement
phase transitions.  An earlier prediction of ref.~\cite{Fodor:2001pe}
suggested that the critical point is approximately at twice as large baryochemical
potentials, suggesting $T_E = 160 \pm 3.5 $ MeV and $\mu_E = 725 \pm 35$ MeV.}
\end{figure}

Various theoretical considerations suggested that at high baryon densities there exists 
a line of first order phase transitions on the QCD phase diagram, and this line ends in a critical end point (CEP), 
where the phase transition changes from a first to  
a second order transition. At even smaller baryochemical potentials (or net baryon densities)
the transition is expected to  become a cross-over, i.e. a rapid increase in the number of degrees of freedom but without
a clear thermodynamical separation of the confined and the deconfined phases. These ideas are illustrated
in Fig.~\ref{f:QCD-phase-diagram}. 

Hence it is of fundamental interest to locate the critical end point of the line of first order QCD phase
transitions: this point acts as a landmark or a sign-post on the QCD phase diagram, and separates the line of 
first order phase transitions from a cross-over region~\cite{stephens:private}.
The next section discusses how to locate, identify and characterize such a critical point
in high energy particle and nuclear physics, based on analogies with similar  experiments
in physical chemistry and on applications of the  theory second of order phase transitions. 

\section{Critical Opalescence: Signature of a  2nd Order Phase Transition}

Before entering  a discussion on how to characterize a critical point, 
let us first discuss a strategy how a critical point is located in statistical physics and
in experiments on second order phase transitions involving ordinary gases and fluids.
This subject will lead us to a new, experimental definition of opacity in high energy heavy ion
physics. In the next section we discuss how to characterize
such a critical point by critical exponents, based on the theory of second order phase transitions in 
statistical physics and the adaptation of these results to heavy ion physics.

\begin{figure}[tb]
\begin{center}
\includegraphics[width=0.30\textwidth]{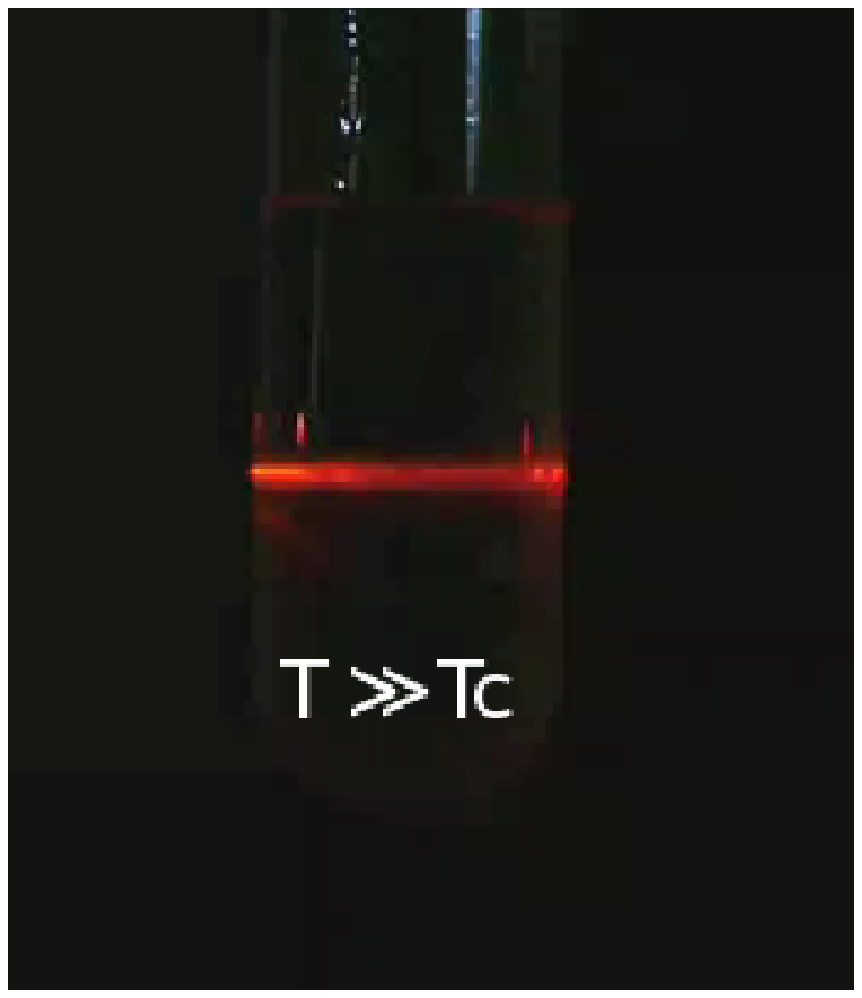}
\includegraphics[width=0.30\textwidth]{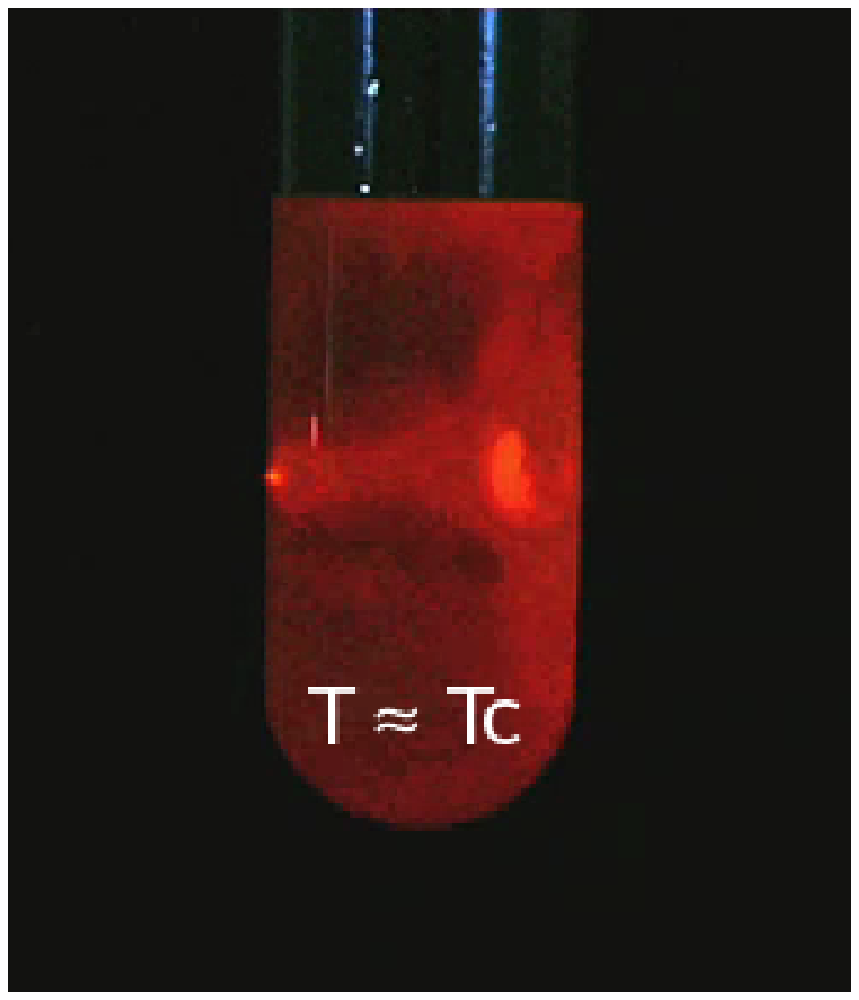}
\includegraphics[width=0.30\textwidth]{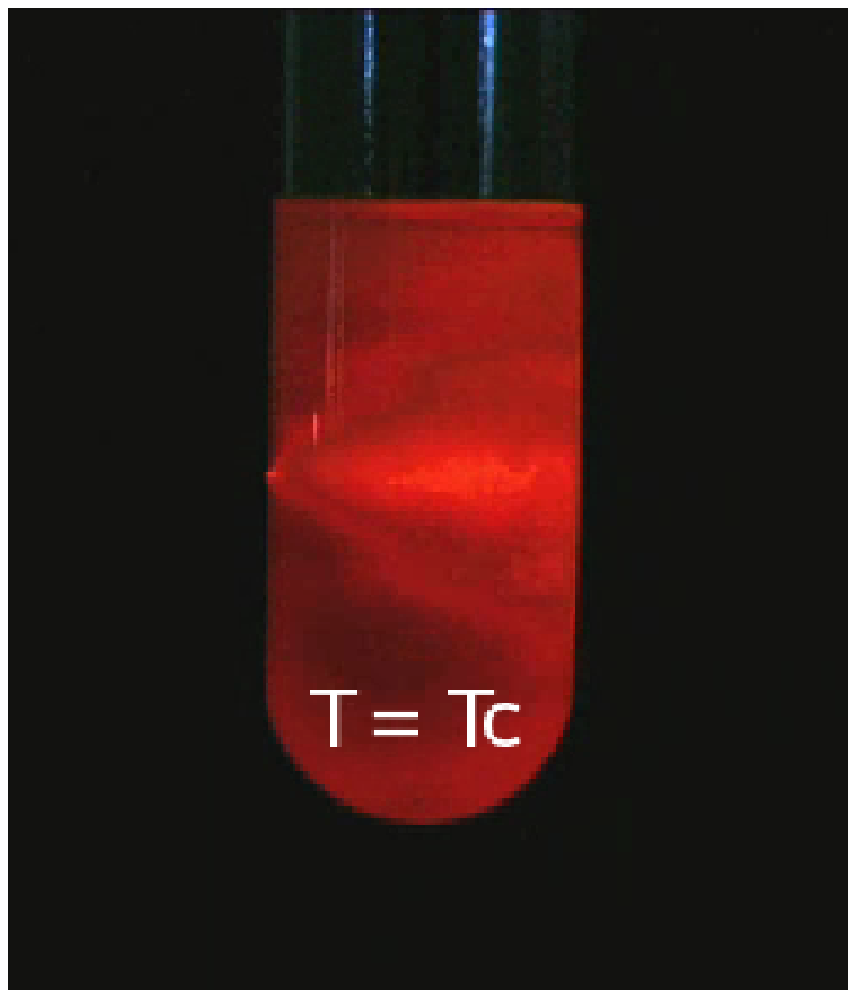}
\end{center}
\caption{%
Locating the critical point in physical chemistry by the phenomenon
called critical opalescence. In a second order phase transition, scattering centers (density fluctuations)
appear on all scales, hence the matter becomes opaque, as indicated by increased scattering of
a penetrating probe, in this case a laser beam. The maximum of the opacity corresponds to the
critical point. Pictures from ref.~\cite{CEP:videos}.
}
\label{f:laser-CEP}
\end{figure}

Critical opalescence is a characteristic phenomenon that arises in second order phase transitions,
as illustrated in Fig.~\ref{f:laser-CEP}. It is commonly demonstrated in binary fluid mixtures,
or at the end point of the line of first order liquid-gas phase transitions. As the critical point is approached, 
the sizes of the gas and the liquid region begin to fluctuate on increasingly large length scales, eventually 
including the size comparable to the wavelength of light. These fluctuations then scatter light and cause the normally
transparent fluid to appear cloudy or opaque. At the critical point, density fluctuations appear on every length scale,
in ordinary fluids these scales include the wavelength of light (from 700 nm to 400 nm) up to as much as a few  mm.
The increased opaqueness of the fluid is a well visible signal of the vicinity of a critical point.

In optics, the opacity $\kappa$ is defined as the rate of absorption or extinction, at which the intensity 
$I$ of a beam of radiation is absorbed or scattered per unit distance along a ray of propagation:
\begin{equation}
\frac{\partial I}{\partial x} = - \kappa I
\end{equation}
The opacity $\kappa$ may depend on the wavelength (or momentum) of the radiation, as well as on the
density, temperature and composition of the medium. For example, air has nearly zero opacity for visible light,
and for radio waves, but it is almost completely opaque for gamma and X rays, and in most of the infrared 
part of the spectrum.  The solution of the above differential equation for constant values of opacity is
\begin{equation}
I = I_0 \exp( - \kappa x)  = I_0 \exp(- x/\lambda) \label{e:opaque}
\end{equation}
where $\lambda = 1/\kappa$ is the so called attenuation length or penetration depth, defined as the distance
where a radiation inside a material falls to $1/e ~\approx 36.7 \%$ of its incoming intensity. 
So a characteristic signature of critical opalescence is a minimum of the attenuation length or
a maximum  value of opacity on many wavelengths where the matter was previously transparent.
Measuring opacity as a function of pressure and temperature, the critical point can be located.
The location of the critical point $(T_c, p_c)$ on the $(T,p)$ diagram is an important characteristics
 of the given material.

\subsection{Opacity in heavy ion collisions}
It is interesting to rewrite the defining relation of optical opacity in a manner that has a straightforward
generalization to heavy ion physics:
\begin{equation}
\kappa = \frac{I(\mbox{\rm generated})-I(\mbox{\rm transmitted})}{I(\mbox{\rm generated})\Delta x} ,
\end{equation}
where $\Delta x$ stands for the distance traveled in the medium.

In heavy ion physics, the nuclear modification factor $R_{AA}$ can be defined as
\begin{equation}
R_{AA}= \frac{I(\mbox{\rm transmitted})}{I(\mbox{\rm generated})} 
 = \frac{I(\mbox{\rm measured})}{I(\mbox{\rm expected})} ,
\end{equation}
where the measured yield is defined as 
\begin{equation}
I(\mbox{\rm measured}) = \frac{1}{N_{event}^{AA}}\frac{d^2N_{AA}}{dy dp_t},
\end{equation}
while for point-like processes like production of high transverse momentum jets the generated
rate is the same yield measured in proton-proton reactions scaled up by $\langle N_{coll}\rangle$,
the number of binary initial nucleon-nucleon collisions, given by a Glauber model calculation:
\begin{equation}
I(\mbox{\rm expected}) = \frac{\langle N_{{coll}}\rangle}{\sigma^{NN}_{{inel}}}\frac{d^2\sigma_{NN}}{dy dp_t}.
\end{equation}
This nuclear modification factor $R_{AA}$ includes a mixture of hot and cold nuclear matter 
effects~\cite{WGH:private},
its most apparent shortcoming is that it measures the change in intensity but does not specify
the distance over which the change of the yield happened. 
Similarly, the nuclear modification factor can be smaller for a matter which has larger opacity, 
if this matter is present only in a short distance 
along the line of propagation of the probe, as compared to a less opaque matter that has
been created in heavy ion collision in a larger volume. 
Opacity as defined in optics, on the other hand, compares the rate
of change of intensities per unit distance. 
Thus for a proper definition, both the nuclear modification factor
and the characteristic length scale of the attenuation has to be included.
Let us consider, for simplicity, a one dimensional symmetric distribution with a characteristic
radius of $R$. Suppose that a pair of back-to-back jets is created inside this distribution and the
near-side jet can escape after covering a distance $r$. Obviously, the far-side jet has to propagate
over a distance of $2 R - r$. Averaging this over the source distribution with the distribution of the
production points $r$ the near side jet travels $\langle r\rangle$ while the far side jet travels
$2 R -\langle r\rangle$ distance in the medium on the average. Averaging over both jets the mean distance
of propagation becomes just $R$, the characteristic radius of the source. 

Such a characteristic source size is measured from two-particle correlations
using various techniques,  named as femtoscopy in general. More specifically, we 
measure the Bose-Einstein correlations  of like-sign pions and from this measurement
in a Gaussian approximation to the source density distribution effective source sizes
are extracted.  These radii are frequently referred to as Hanbury Brown - Twiss or
HBT radii to honor the two radio astronomers, who invented a similar technique for
photons to measure the angular diameter of main sequence stars.
These HBT radii can be (and apparently are~\cite{Adams:2003ra}
) measured as a function of the angle with respect to the reaction plane.
For a small attenuation or $1-R_{AA} \approx \epsilon$, we may define the measurable opacity as
$
\kappa_\epsilon = \frac{1 - R_{AA}}{R_{HBT}} 
$
where we divide the relative attenuation of intensity by $R_{HBT}$,
the characteristic length scale of the fireball measured (in principle) along
the line of propagation.
(The application of Bose-Einstein or Hanbury Brown - Twiss correlation techniques 
are detailed in subsequent sections.)

It is important to note that in high energy heavy ion reactions in the $p_t > 4$ GeV
kinematic range we are measuring not tiny absorbtions, but quite significant values, so
the thickness of the absorber or the fraction of the absorbed intensity cannot be considered
small, $1  - R_{AA} \approx 0.8$. Hence   we have to solve eq. (\ref{e:opaque}) to get
a more precise definition, keeping in mind the interpretation of the nuclear absorbtion factor
as $R_{AA}=I/I_0$. Hence for arbitrary thickness and nuclear modification factor, the
optical opacity should properly be defined as
\begin{equation}
\kappa = - \frac{ \ln(R_{AA})}{R_{HBT}} \label{e:kappa}
\end{equation}

The opacity  $\kappa$ defined above  may depend in principle on  system size, colliding energy, centrality,
and angular direction with respect to the reaction plane, through $R_{AA}$ as well as $R_{HBT}$.
As the average HBT radius as well as $R_{AA}$ are both positive, the above quantity is well defined. 
Using the above definition,
we can compare the opacity of different media that might be available in differing amounts.
This definition has several other advantages: for $0 < R_{AA} < 1$ the opacity is positive as the HBT radius
is always positive.  In case of no nuclear modification, $R_{AA} = 1$ and $\kappa = 0$, the opacity vanishes, indicating
that the medium is fully transparent. On the other hand, if $R_{AA} > 1$, the opacity becomes negative, indicating that the
matter is not absorbing but in fact emitting extra radiation. 
However, the main advantage of the above definition
is that it defines opacity as a modification of spectra per unit length along the direction of
the propagation of  a penetrating probe.
Thus it characterizes a local property of the medium, and not an integral property. Local and not integral properties
are relevant in a search for a phase transition: it may happen that the medium created in central $Cu+Cu$ collisions
has less absorption or has a larger $R_{AA}$ than that of a mid-central $Au+Au$ collision simply because
the bulk amount of the medium is larger in mid-central $Au+Au$ collisions. Comparing to  
the HBT radius is necessary to correct opacity for the bulk effect. 

It is even more intuitive to introduce the inverse of opacity, as the attenuation length:
\begin{equation}
\lambda \, = \, \frac{1}{\kappa} \, = \, - \frac{R_{HBT}}{ \ln( R_{AA})}. \label{e:lambda}
\end{equation}
This measures the length over which the intensity of radiation is attenuated by a factor of $1/e$.
In case of some nuclear absorption, $0 < R_{AA} < 1$, and  $0 < \lambda < \infty$.
For no nuclear modification and $R_{AA} = 1$, the attenuation length diverges and it becomes negative for
$R_{AA} > 1$, as the medium enhances the radiation. Thus a negative $\lambda$ means
that the intensity decreases in the direction opposite to that of the propagation.

So in principle, one has to infer the actual distance that the probe had to cross before departing towards the detectors.
It is well known that the size of nuclei change on the average as $A^{1/3}$, which in a collision
would then be modified to an average $N_{{part}}^{1/3}$  
dependence.  The freeze-out volume (more precisely, the size of the region of homogeneity) of 
charged pion emission was measured in 200 GeV Au+Au 
collisions using Bose-Einstein or HBT techniques as a function of centrality~\cite{Adler:2004rq}. 
The Gaussian HBT radii were found to scale as 
$R_i = p_0 + p_1 N_{{part}}^{1/3}$, where $p_0 \approx 0.83 \pm 0.25 $ fm and $p_1 \approx 0.54 \pm 
0.05 $ fm and $i$ = (long, out, side) stands for the three principal directions with
respect to the beam and the mean transverse momentum of the particle pair.
Note that ref. ~\cite{Adler:2004rq} 
fitted the  centrality dependence of each 
HBT radius components separately, but let me point out that the fit parameters 
in the different directions were found, within one standard deviation, to have the same values.
Thus their average value was quoted above. Experimentally,
$R_{{out}}\approx
R_{{side}}\approx 
R_{{long}} \approx R_{HBT} $,  due to the effective spherical symmetry of the pion emitting source.
Note also that penetrating probes (like high $p_t$ jets, leptons or gammas) can be radiated, in principle,
throughout the time evolution of the expanding fireball created in high energy heavy ion collisions.
So the distance that is covered by these probes in the medium is not a static distance, but a length scale 
of the expanding source that is covered by the jets before they punch through.
Few exact results are known for the time evolution of such mini-bangs or exploding fireballs
of heavy ion reactions, but two points are quite remarkable: in a non-relativistic approximation
it was found that although each principle directions of the expanding fireball diverges to infinity,
the HBT radii or lengths of homogeneity tend to a time and direction independent constant,
so the HBT volume becomes spherical~\cite{Csorgo:2002kt}; 
second that experimentally such a spherically symmetric HBT volume
is seen in pion-correlation data~\cite{Adler:2004rq}; third that the HBT volume 
-- just like the initial volume -- is found to scale with $N_{part}^{1/3}$.
Based on these three observations we may expect that the distance covered by the probe
in the medium is indeed proportional to $N_{part}^{1/3}$, hence scaling $(1-R_{AA})$ by the
HBT radii that are also proportional to $N_{part}^{1/3}$ gives the centrality dependence
of the opacity $\kappa$ correctly; furthermore the offset parameter $p_0 > 0$ guarantees that
even in the most peripheral collisions the initial volume that has to be propagated through
by the high $p_t$ probes is non-vanishing.
Such expansion effects may indeed introduce certain absolute calibration error on the opacity $\kappa$
that is difficult to control. However, the relative error on the centrality dependence of the
 opacities $\kappa$ and in particular its minimum or maximum structure  is not affected
by these expansion effects, as both the initial and the final volumes scale proportional 
to $N_{part}^{1/3}$. 

In principle, both the effective source size  $R_{HBT}$,
and the nuclear modification factor $R_{AA}$ may change with respect to the reaction plane even in events
with the same colliding system, bombarding energy and centrality. Then opacity should be defined by 
the above ratio evaluated as a function of angle with respect to the reaction plane. 
Although both $R_{AA}$ and $R_{HBT}$ may change with respect to the orientation 
of the reaction plane, the optical opacity is expected to be independent of the selected directions
in a fireball that  consists of one given type of matter. 

As a test of the proposed method, and 
for simplicity, let me use here an average
HBT radius as $R_{HBT} = (R_{{out}}+ R_{{side}}+ R_{{long}})/3$
and also nuclear modification factors of neutral pions  at $p_t = 4.75$ GeV in  PHENIX 
to evaluate the opacity and the attenuation lengths in $\sqrt{s_{NN}}=200$ GeV Au+Au collisions, as
a function of centrality. 
The results are shown in 
Table~\ref{table:opacity}.  We find that changing centrality 
from 50 - 60\% to  0-5 \%, the opacity is increased by substantially, by about a factor of 2 (a 4 $\sigma$ effect), 
while the nuclear modification factor decreases by more than a factor of 4 from 0.72 to 0.19.  Thus
nearly half of the suppression increase originates from the increased amount of the
hot and dense, strongly interacting matter created in more central collisions,
while the attenuation length is also decreased by a factor of $2$ in these collisions. 

%
\begin{table}[ht]
\centering
\begin{tabular}{l c c c c c}
\hline\hline
Centrality &  0-5 \% &  20-30 \% & 30-40 \% & 40-50 \% & 50-60 \%\\ [0.2ex]
\hline
Opacity $\kappa$ (fm$^{-1}$) & 0.35  $\pm$ 0.04 & 0.27 $\pm$ 0.03 & 0.26 $\pm$ 0.04 & 0.12 $\pm$ 0.02 & 0.15 $\pm $ 0.05 \\
 [0.2ex]
Attenuation $\lambda$ (fm) & 2.9  $\pm$ 0.3 & 3.7  $\pm$ 0.4 & 3.8  $\pm$ 0.6  & 8.1  $\pm$ 1.5 & 6.5 $\pm$ 2.0 \\ [0.2ex]
\hline 
\hline
\end{tabular}
\label{table:opacity}
\caption{
Examples of opacities  $\kappa $ 
and attenuation lengths $\lambda $ 
in $\sqrt{s_{NN}}= 200$ GeV Au+Au reactions,
evaluated from nuclear modification factors measured at $p_t = 4.75 $ GeV/c in ref.~\cite{Adare:2008qa}
and using HBT radii of ref.~\cite{Adler:2004rq},  averaged over both  directions  and charge combinations,
at the same centrality class as  $R_{AA}$. }
\end{table}

For a precise absolute calibration of the opacity $\kappa$, additional information is needed
both in the soft and in the hard transverse momentum sector. To measure the change of 
intensities at high transverse momentum, we have to know precisely what the
initial energy of the jet was, before it became attenuated. This is possible to learn, if we trigger
 on $q+\overline{q} \rightarrow \gamma + g$  or $q + g \rightarrow q + \gamma$
events with  high $p_t$   $\gamma$-s,
 and measure the energy of the high transverse momentum photon precisely,
because $\gamma$'s are interacting only electromagnetically 
so they are not attenuated in nuclear matter.
With the help of such processes, the energy and momentum of the momentum balancing
hadron jet, or the initial intensity can be measured precisely. However, about two orders of magnitude more
events are necessary for such a study, than for a usual $R_{AA}$ measurement, due to the relatively 
small cross-sections and large $\gamma$ background from $\pi^0$ and other hadronic decays.
Similarly, the time evolution of the size of the  expanding fireball could in principle be measured by
soft $\gamma + \gamma$ correlations, as these probes penetrate the hadronic fireball too;
but up to now no such studies were published in the RHIC program, due to difficulties
in background rejection and the need for extremely large statistic. We leave these topics
as future challenges to the RHIC experiments.

\section{Critical Exponents and the Universality Class of the Critical Point}
Phase transitions that involve latent heat are called first order phase transitions. 
During such phase transitions, the system releases or absorbs a fixed amount of energy while its
temperature stays constant as heat is removed or added to the system. In this period, two different
phases with similar temperatures and pressures but with two different internal energy densities co-exist,
 only their volume fraction changes, until one of them
reaches 100 \% and the other disappears. This phase coexistence region is frequently and incorrectly 
referred to as "mixed phase" although correctly this does not correspond to a single (but mixed) phase,
instead it corresponds to a mixture of two, locally different, but co-existing phases.
At very high baryon densities, the rehadronization phase transition of a quark-gluon plasma is believed
to be such a first order phase transition, during which a specific amount of latent heat is released.
As the energy density of the matter can change at a constant pressure,
the speed of sound typically vanishes during such a first order phase transition.
At the critical point $(T_c,\mu_c)$ the latent heat disappears,
and the order of the phase transition changes from a first order to a second order one.
In the vicinity of the critical point, the difference between the gaseous and the liquid phase becomes almost non-existent.  Fluctuations appear on all possible wavelengths and are signaled by critical opalescence, or the decreased
attenuation length at all possible wavelengths.

As the characteristic sizes disappear from the physical description of second order phase transitions at the
critical point, power-law distributions emerge and characterize the physical quantities and the exponents
of these  power-laws, called the critical
exponents become the  relevant, physically important observables.
Frequently, a reduced temperature  is introduced:
\begin{equation}
 t = \frac{T-T_c}{T_c}
\end{equation}
Along the line of phase transitions, a small change in temperature corresponds to a small change in
baryochemical potential, $\Delta T \propto \Delta \mu_B$, so the power law exponents will be similar if the fluctuations
are studied in terms of deviations from the critical temperature, or if $t$ is re-defined 
as dimensionless deviation from the critical value of the baryochemical potential:
$ t^\prime = \frac{\mu_B-\mu_{B,c}}{\mu_{B,c}}.$

To characterize the phases, an order parameter $|\phi|$ is introduced, which is zero before and non-zero after the
phase transition, conventionally so that $|\phi| > 0$ corresponds to $T < T_c$. 
In case of QCD phase transitions, a theoretical order parameter can be the
expectation value of the quark condensate, $|\phi| = \langle q\overline{q} \rangle $. 
Experimentally one can, for example, define an order parameter, 
that measures the violation of  the quark number scaling of elliptic flow of identified kaons
$(v_2^K)$ and identified protons $(v_2^p)$ , and use it as
a phenomenological order parameter: $|\phi^\prime| = \int d\mu(w) 
\left[\frac{1}{2} v_2^K(\frac{w}{2}) - \frac{1}{3} v_2^p(\frac{w}{3})\right]^2$. 
For the definition of the universal scaling variable $w$ and that of the elliptic flow $v_2$
see ref.~\cite{Csanad:2005gv}, but note that at mid-rapidity $w \propto (m_t - m)/T_*$,
where $m_t-m = KE_t$ is the transverse kinetic energy and $T_*$ stands for the slope
parameter of the single particle spectra.  The limits of integration are given to the
upper kinematic limit of the applicability of a hydrodynamical picture, 
typically $p_t \approx 1.5 - 2$ GeV. The measure $d\mu(w)$ may be chosen appropriately,
for the simplest choice I recommend to try $d\mu(w) = dw$, but of course other measures
like $d\mu(1;w) = \exp(-w) dw $ or $d\mu(\lambda;w) = \exp(-\lambda w) dw $ etc
are also possible. It is physically important to remove data points that are not
from the hydrodynamic kinematic range, i.e. close to the target or projectile rapidity
or have large $p_t > 2$ GeV.
The integral above has a vanishing argument
if the elliptic flow corresponds to a perfect fluid of quarks, othervise it is positive,
hence it can be used as an order parameter, chosing any form of the above measures.
Of course similar order parameters can be defined for other particle pairs. 
Typically a baryon and a meson should be paired, to utilize their different number of
valence quarks. I recommend to leave out pions, as their light mass (would-be Goldstone boson nature)
and the unknown confinement of two heavy valence quarks into a single light pion
makes their elliptic flow connection to the quark elliptic flow rather subtle and
theoretically challenging.

The new order parameter defined above is vanishing in a fluid of quarks and anti-quarks
(better known as a strongly interactive quark-gluon plasma phase),
while it is expected to become non-vanishing in a hadronic  phase,
i.e. when the quark number scaling of the elliptic flow is broken.

We introduce below the six characteristic  critical exponents of the second order phase transitions,
$\alpha$, $\beta$, $\gamma$, $\delta$, $\eta$ and $\nu$
following the lines of ref.~\cite{Rajagopal:1992qz}.

The heat capacity at constant volume $V$ or pressure $p$ is defined as
$C = T \frac{\partial S}{\partial T}|$ where the derivative is taken at constant values of  $V$ or $p$, respectively.
Refs.~\cite{Stodolsky:1995ds,Shuryak:1997yj} suggested that event-by-event fluctuations of 
the temperature $T$ be used to measure
the heat capacity at constant volume as
\begin{equation}
\frac{1}{C_V} =  \frac{(\Delta T)^2}{T}.
\end{equation}
A caveat is if the parameter $T$ in this equation should be obtained from fits to the single particle spectra
($T_{kin}$, the kinetic freeze-out temperature) or from fits to particle abundances ($T_{chem}$ the chemical freeze-out temperature).
Given the expectation that $T$ in the above equation refers to temperatures in the vicinity of the critical point,
 I recommend to use temperature measures that yield values closer to the line of the phase transitions,  
e.g. $T=T_{chem}$, because $T_{chem} \ge T_{kin}$. 
Note however that both temperatures may have a spatial  distribution even in a single event,
so this measure can be considered only as an approximate one.
In a second order phase transition, the critical exponent $\alpha$ measures the power-law behavior of this  heat capacity
near the critical point:
\begin{equation}
C(t) \simeq |t|^{-\alpha},
\end{equation}
where $\alpha$ may have both negative and positive values, as for example in the case of onset of superfluidity in $^4$He,
or for the case of the 3d Ising model, respectively.
If the baryochemical potential tends to  vanishing values as the critical point is approached, $\mu_B
= a (T-T_c)^b$ for some positive constants $a$ and $b$, 
the specific heat capacity of baryons obeys the above law with $\alpha = 2$, as shown
in ref.~\cite{Mekjian:2006dw}.

Near but below the critical point, the temperature dependence of the order parameter 
is given by the critical exponent $\beta$:
\begin{equation}
|\phi(t)| \simeq |t|^{\beta}, \qquad \mbox{\rm for}\quad t < 0.
\end{equation}

The correlation function of the order parameter is defined as
\begin{equation}
G(r,t) = \langle \phi(r,t) \phi(0,t) \rangle - \langle \phi(0,t)\rangle^2,
\end{equation}
where $r$ is a measure of relative coordinates.
At large distances and for $ t \ne 0$,
\begin{equation}
G(r,t) \rightarrow  \frac{A}{|r|} \exp(-|r|/\zeta),
\end{equation}
where $A = A(t) $ and the correlation length $\zeta = \zeta(t)$ are both temperature dependent, 
but both are independent of relative separations.
The susceptibility exponent $\gamma$ is defined as
\begin{equation}
\int d^3 r G(r,t) \propto |t|^{-\gamma}. 
\end{equation}
The exponent $\delta$ in the 3d Ising model is related to a small external field $H$ that breaks the
$O(4)$ symmetry of the model explicitly. In terms of QCD, such a symmetry breaking term 
is related to the small but non-vanishing
value of the masses of light quarks, $H \propto m_q = m_u = m_d$.
In the presence of an external field, the order parameter is not vanishing at the critical temperature, but
scales with the magnitude of the symmetry breaking terms as
\begin{equation}
\langle \phi(t=0, H\rightarrow 0)\rangle \propto |H|^{1/\delta}. 
\end{equation}
Two very important critical exponents are $\nu$, the exponent of the correlation length $\zeta$, defined as
\begin{equation}
\zeta \propto |t|^{-\nu}, 
\end{equation}
and $\eta$, the exponent  of the Fourier-transformed correlation function 
in the critical point ($t=0$), defined as
\begin{equation}
\tilde G(k\rightarrow 0) \propto |k|^{-2 + \eta}. 
\end{equation}
Out of these six critical exponents only two are independent. The remaining four are related by  
the theory of second order phase transitions as
\begin{eqnarray}
\alpha & = & 2 - d \nu,\cr
\beta & = & \frac{\nu}{2} (d -2 + \eta), \cr
\gamma & = & (2 - \eta) \nu, \cr
\delta & = & \frac{d+2 - \eta}{d- 2 +\eta},
\end{eqnarray}
where $d$ stands for the number of dimensions, $d= 3$ for heavy ion collisions.
See ref.~\cite{Rajagopal:1992qz} and references therein for more details on how to obtain
these relationships from the assumption that the free energy at the critical point is described by
scale invariant, generalized homogeneous functions.

It is quite remarkable that second order phase transitions of apparently very different
systems are often characterized by the same set of critical exponents. This phenomenon is
called universality. For example, critical exponents at the critical point of the liquid-gas phase
transition have been found to be independent of the chemical composition of the fluid.
Furthermore, they are exactly the same as the critical exponents of the ferromagnetic phase 
transition in uniaxial magnets, which is the same as that of the 3 dimensional Ising model. 
Such systems are then referred to as being in the same {\it 
universality class}. The renormalization group theory of second order phase transitions
states that the thermodynamic properties of a system near a critical point
depend only on a small number of features such as the dimensionality of the system and 
the symmetry of the interactions, but are insensitive to the underlying microscopic properties. 
That is why it will be so interesting to test, if the perfect liquid of quarks, created in Au+Au collisions
at RHIC, features a critical point or not. It is the universality class of QCD
that characterizes matter at this point. This universality class is the key
observable that experiments will have to determine, if indeed such a critical point of QCD is found.
Rajagopal and Wilczek predicted already in 1992~\cite{Rajagopal:1992qz} that the universality class of QCD will be
that of the 3d Ising model, where the critical exponents are measured as
\begin{eqnarray}
	\eta &=& 0.03 \pm 0.01,\cr
	\nu &=& 0.73 \pm 0.02,
\end{eqnarray}
hence the remaining exponents are given as
\begin{eqnarray}
	\alpha &=& - 0.19 \pm 0.06,\cr
	\beta &=& 0.38 \pm 0.01,\cr
	\gamma &=& 1.44 \pm 0.04,\cr
	\delta &=& 4.82 \pm 0.05.
\end{eqnarray}

We have conjectured that the violent environment of a heavy ion collision generates
random fields that interact with the  expanding medium and hence change the
universality class of the 3d Ising model to that of the random field 3d Ising model~\cite{Csorgo:2005it}.
In particular, due to this expected noise, the extremely low correlation exponent
$\eta$ increases,  and the critical exponents are given ~\cite{Rieger:1995it} as
\begin{eqnarray}
	\eta &=& 0.50 \pm 0.05,\cr
	\nu &=& 1.1  \pm 0.2.
\end{eqnarray}
hence the remaining exponents are given as
\begin{eqnarray}
	\alpha &=& - 1.3 \pm 0.6,\cr
	\beta &=& 0.6 \pm 0.1,\cr
	\gamma &=& 2.2 \pm 0.4,\cr
	\delta &=& 4.7 \pm 0.3 .
\end{eqnarray}
Note that the changes are most significant in the exponents $\eta$ and $\nu$ and these are
the exponents that are most  directly measurable, as will be discussed below.

\section{Soft correlations for measuring critical exponents of the correlation function}
In this subsection, we discuss how to measure 
the critical exponent of the correlation function $\eta$, based on a shape analysis of
the Bose-Einstein or HBT correlation functions.

Bose-Einstein correlation functions
in high energy physics for L\'{e}vy stable sources were introduced in  
refs.~\cite{Csorgo:2003uv,Csorgo:2004ch}. These papers
and ref.~\cite{Csorgo:2005it} 
are recited in this subsection,
to point out how to measure the critical exponent of the correlation
function in the vicinity of a critical point.
	
Bialas considered a distribution of Gaussian coordinate space distributions,
where the Gaussian radius parameter has a power-law distribution~\cite{Bialas:1992ca}, 
and related the  Bose-Einstein correlations of such sources to scale-invariant multiplicity 
fluctuations called intermittency.  Brax and Peschanski introduced L\'{e}vy
distributions to multiparticle production in high energy physics~\cite{Brax:1990jv},
but in momentum space.  They suggested to use the measured values
of the L\'evy index of stability to signal quark gluon plasma production in heavy ion physics.
Here we follow ref.~\cite{Csorgo:2005it} 
in reconnecting these apparently different topics.

\subsection{Bose-Einstein correlations for L\'evy stable source distributions }
The Bose-Einstein correlation function of identical boson pairs
can be defined with the help of the corresponding two-particle 
and single-particle invariant momentum distributions: 
\begin{equation}
C_2({\mathbf k}_1,{\mathbf k}_2) = 
\frac{N_2({\mathbf k}_1,{\mathbf k}_2)}
{N_1({\mathbf k}_1)\, N_1({\mathbf k}_2)}.  \label{e:cdef}
\end{equation}
If the long-range correlations can be either neglected or be corrected for,
and the short-range correlations are dominated by Bose-Einstein
correlations, this two-particle Bose-Einstein correlation function
is related to the Fourier-transformed source distribution.

For clarity, let us consider a one-dimensional and static source,
where the space-time distribution $f(x)$ and the momentum space
distribution  $g(k)$ are factorized:
\begin{equation}
S(x,k) =  f(x) \, g(k).
\end{equation}
The corresponding Bose-Einstein correlation function is
\begin{equation}
C_2(k_1,k_2) = 1 + |\tilde f(q)|^2,
\end{equation}
where the Fourier transformed source density (often referred to as the
{\it characteristic function}) and the relative momentum are defined as
\begin{equation}
\tilde f(q)  = \int  \mbox{\rm d}x \, \exp(i q x) \,f(x),\qquad\quad 
q  =  k_1 - k_2 .\label{e:fourier}
\end{equation}
During a second order phase transition, where fluctuations appear
on all possible scales with a power-law tailed distribution, the final position
of a particle is given by a large number of position shifts,
hence $f(x)$, the distribution of the final position $x$ is obtained as  
\begin{equation}
x = \sum_{i=1}^n x_i, \qquad\qquad
f(x) = \int \Pi_{i=1}^n \mathrm{d} x_i\, \Pi_{j=1}^n f_j(x_j)\,
\delta(x - \sum_{k=1}^n x_k ).
\end{equation}
A system at a second order phase transition becomes invariant under 
a renormalization group transformation. If we consider particle production
near to such a scale invariant point, we may expect that adding an $(n+1)$th
random shift does not change (after rescaling) the overall source distribution. 
Thus at a second order phase transition we expect  $f(x)$ to be stable
for convolution. 

In probability theory, Central Limit Theorems state
that under certain conditions, the distribution of the sum of
large number of random variables converges (for $n \rightarrow \infty$)
to a limit distribution.
If the elementary steps obey a ``normal" elementary distribution
that has a finite mean and variance, then  the limit distribution of their sum 
can be proven to be a Gaussian.
However, in the vicinity of a second order phase transition point, 
fluctuations appear on all scales and the variance of the elementary process diverges. 
In this case, the probability distribution will deviate from a Gaussian, 
in particular it develops power-law tails, but it still may be stable for convolutions. 

\begin{figure}[t]
\begin{center}
\includegraphics[width=0.6\textwidth]{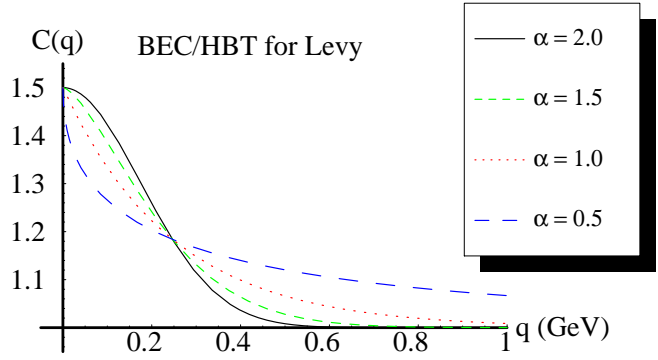}
\end{center}
\caption{%
Bose-Einstein or HBT correlation functions for L\'evy stable source
distributions for various indices of stability $\alphal$. At a second order
QCD phase transition, the index of stability becomes one of the critical exponents
of the phase transition, namely the exponent of the correlation function, $\eta = \alphal$.
Note that on this plot, $\alpha$ stands for the L\'evy index of stability, $\alphal$,
and it is not to be confused with the critical exponent of the heat capacity, which is denoted by the
same symbol in the body of the manuscript.
From Ref.~\cite{Csorgo:2005it}.
}
\label{f:Levy-HBT}
\end{figure}

Stable distributions are limit distributions
that occur in Generalized Central Limit theorems. The study
of such distributions was started by the French mathematician 
Paul L\'{e}vy in the 1920's.
Stable distributions can be given simply in terms of
their characteristic functions, as the Fourier transform of
a convolution is a product of the Fourier-transforms,
\begin{equation}
\tilde f(q) = \prod_{i=1}^n \tilde f_i(q) .
\end{equation}
The characteristic function of univariate and symmetric stable distributions is
\begin{equation}
\tilde f(q)=\exp\left( i q \delta -|\gamma q|^\alphal\right), \label{e:fqs}
\end{equation}
where the support of the density function $f(x)$ is $(-\infty,\infty)$.
Deep mathematical results imply that the index of stability, $\alphal$,
satisfies the inequality $0 < \alphal \le 2$,
so that the source (coordinate) distribution be always positive.
These L\'{e}vy distributions are indeed stable under convolutions,
in the sense of the following relations:
\begin{eqnarray}
\tilde f_i(q) & = & \exp\left( i q \delta_i -|\gamma_i q|^\alphal\right), \qquad
\prod_{i=1}^n \tilde f_i(q) \, =  \,
\exp\left( i q \delta -|\gamma q|^\alphal\right) ,\\
\gamma^\alphal & = &  \sum_{i=1}^n \gamma_i^\alphal,
\quad\qquad\qquad\qquad
\delta \, = \, \sum_{i=1}^n \delta_i. \label{e:gami}
\end{eqnarray}

Thus the Bose-Einstein correlation functions for uni-variate,
symmetric stable distributions (after a core-halo correction, and a re-scaling) read as
\begin{equation}
C(q;\alphal) = 1 + \lambda \exp\left(-|q R|^\alphal\right).  \label{e:BEC-Levy}
\end{equation}
See refs.~\cite{Csorgo:2003uv} and ~\cite{Csorgo:2004ch} for further examples and details.

\subsection{Bose-Einstein correlations at a  second order QCD phase transition}
	At the CEP, the second order phase transition is characterized by
	the fixed point of the renormalization group transformations.
	In a quark-gluon plasma, the vacuum expectation value of the
	quark condensate $ \phi = \langle \overline q \, q\rangle$ vanishes,
	while in the hadronic phase, this vacuum expectation value becomes
	non-zero.  The correlation function of the order parameter is defined as
	$\rho(r) = \langle \phi(r) \phi(0) \rangle - \langle \phi \rangle^2 $
	and measures the spatial correlation between pions.
	At the CEP, this correlation function decays as
\begin{equation}
	\rho(r) \propto r^{- (d - 2 + \eta)} ,
\end{equation}
a power-law. The parameter $\eta$ is called as the exponent of the correlation function.

\begin{figure}[t]
\begin{center}
\includegraphics[width=0.6\textwidth]{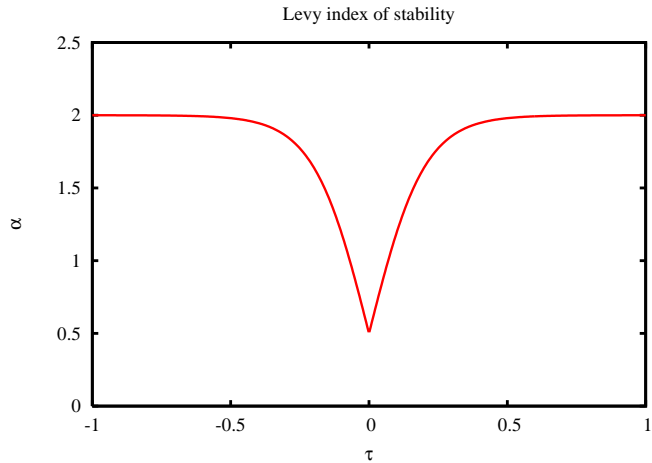}
\end{center}
\caption{%
To locate and characterize the QCD critical point (the endpoint of  the line
of the assumed first order QCD transitions at high baryon densities) 
the L\'evy index of stability $\alpha \equiv \alphal $ is plotted as a function of
reduced temperature, the dimensionless variable $\tau$.
From Ref.~\cite{Csorgo:2005it}.
}
\label{f:Levy-alpha-T}
\end{figure}

For L\'{e}vy stable sources, corresponding to an anomalous diffusion
with large fluctuations in coordinate space, the correlation between the
initial and actual positions decays also as a power-law, where the
exponent is given by the L\'{e}vy index of stability $\alphal$ as
\begin{equation}
	\rho(r) \propto r^{- (1 + \alphal)} .
\end{equation}
As we are considering a QCD phase transition in a $d=3$ three-dimensional coordinate space
we find that the correlation exponent equals to the L\'{e}vy index of stability,
$
	\eta = \alphal .
$
Fig.~\ref{f:Levy-HBT}
indicates that the change in the shape of the correlation function
is rather significant, if $\alphal$ decreases from its Gaussian value of 2
to 0.5, its characteristic value conjectured from the random field 3d Ising model.
Fig.~\ref{f:Levy-alpha-T}
illustrates how this shape parameter of the Bose-Einstein correlation
function may depend on the reduced temperature near the critical point.
Hence the {\it shape} parameter of the two-particle
Bose-Einstein or HBT correlation functions
can be used to determine if the pions
are emitted from the neighborhood of the critical end point of the
QCD phase diagram~\cite{Csorgo:2005it}. Unfortunately, the  
bombarding energy or the centrality of this shape parameter
has not yet been determined in heavy ion collisions.

It is recommended that such a sharp drop of the index of stability 
$\alphal$ is searched first as a function of the beam energy,
the size of the colliding nuclei and the centrality of events, 
by measuring the excitation function of
the L\'evy index of stability in the neighborhood of the maximum of
opacity $\kappa$.  
Close to the critical point, where such a drop in the L\'evy index
of stability  is already observed, 
the collected events could be further separated to several sub-classes based
on the event-by-event values of $T_{chem}$, the chemical freeze-out
temperature, fitted to the chemical composition of the given event.
This way, the dependence on $\tau = (T-T_c)/T_c$ of observables like the critical exponent of
the correlation function or that of the correlation length
can be measured.

\subsection{Negative binomial multiplicity distributions and the exponent of the correlation length}
In this subsection we discuss how to measure 
the critical exponent of the correlation length $\eta$. 
An indirect method to measure a quantity proportional to the 
correlation length has already been published by PHENIX in 
ref.~\cite{Adler:2007fj}. Observing that the Bjorken energy density estimate,
hence the expected initial temperature, scales as a monotonous function of the
number of participants, that paper studied the centrality dependence of
the multiplicity distributions in $\sqrt{s_{NN}} = 200$ GeV Au+Au collisions at RHIC.
In various intervals of pseudo-rapidity 
$\eta = 0.5\log\left(\frac{|p| + p_z}{|p| - p_z}\right)$, 
the multiplicity distribution 
was fitted by a negative binomial distribution (NBD) 
\begin{equation}
P_n = \frac{\Gamma(k+n)}{n! \, \Gamma(k)} 
	\left(\frac{\langle n\rangle }{k}\right)^{n}
	\left(1+\frac{\langle n\rangle }{k}\right)^{-k - n} ,
\end{equation}
that has two fit parameters,
the mean multiplicity $\langle n\rangle$ and the width parameter $k$. This parameter is 
related to the second scaled factorial moment of the multiplicity distribution,
\begin{equation}
\frac{1}{k} = \frac{\langle n(n-1)\rangle}{\langle n\rangle^2} - 1.
\end{equation}
Data were tabulated in one dimension, as a function of the pseudo-rapidity.
For sufficiently large pseudo-rapidity intervals, where the correlation
length is smaller than the size of the window, $\zeta << \delta\eta$,
the width parameter is proportional to the correlation length:
\begin{equation}
\frac{1}{k} = 2 a \frac{\zeta}{\delta\eta} + b
\end{equation}
and using this property the product $a \zeta$ has been determined.
Most interestingly, an indication was found for a non-monotonous
behavior of the correlation length at centralities corresponding 
to about 90 participating nucleons~\cite{Adler:2007fj},
in the 35\% - 45\% centrality class. 

However, other observables including opacities of
Table~\ref{table:opacity} 
indicate no signals for a non-monotonic behavior
in the vicinity of this  centrality class.
Clearly, more detailed studies are necessary,
the opacity estimates of Table~\ref{table:opacity} 
for example depend on the assumption that the 
Bose-Einstein correlation function can be fitted with
a Gaussian form, hence a Gaussian $R_{HBT}$ is
a good measure of the length-scale over which the attenuation happened.  
However, such an assumption is clearly not adequate
when looking for a critical point. A possible decrease of the
shape parameter $\alphal$ of the correlation function
at this centralities should be investigated together with
the question if a L\'evy fit to the Bose-Einstein
correlation function would yield an improved description
of Bose-Einstein correlation data and if the
extracted  scale parameters $R_{\mbox{\tiny\it L\'evy}}$ would be significantly
different from the Gaussian radius parameters.

Similarly to the suggested study of the exponent,
a large sample of events, collected close to the interesting centrality class
of 25- 35 \% , need to be separated to several sub-classes based
on the event-by-event fluctuating values of $T_{chem}$, the chemical freeze-out
temperature, given by  the hadro-chemical composition of each event separately.
This way, the dependence on $\tau = (T-T_c)/T_c$ of the correlation length and
that of the opacity could  be measured, to see if
indeed in $\sqrt{s_{NN}} = 200 $ GeV Au+Au collisions a critical phenomenon
at 25-35 \% centralities is approached, or not.


\section{Characterizing a second order phase transition in QCD: a strategy}
The following straightforward strategy is suggested to locate and characterize experimentally 
the properties of a second order QCD phase transition: 

a) Locate experimentally the critical end point (CEP) of the line of 
1st order phase transitions by searching for critical opalescence,
i.e. looking  for the maximum of opacity, or the minimum of attenuation length,
as a function of centrality, bombarding energy and size of the colliding nuclei in a broad,
high transverse momentum range.

b) Determine the chemical freeze-out temperature $T$ on an event-by-event basis
for an event selection that corresponds to the maximum of opacity.
Check that various observables (heat capacity, correlation length) indeed
indicate a power-law behavior as a function of $t = |T-T_c|/T_c$ simultaneously for the 
event class selected in point a). 

c) Determine the critical exponents for at least two observables. Use theoretical relations
to determine the remaining critical exponents, so that all the six critical exponents be given.
It is recommended to try to measure the critical exponent of the correlation function $\eta$ as
it is defined in the momentum-space, and the power-law shape shows up as a function of the relative momentum 
and not as a function of $t$, hence this exponent is easy to measure. The correlation length can be
determined from the multiplicity distribution using the negative binomial distribution, and
the exponent of the correlation length $\nu$ can be obtained from event-by-event determination
of both the correlation length and the chemical freeze-out temperature $T$.

d) Determine the universality class of QCD from the two measured critical exponents. 

e) Cross-check the result by measuring additional 
critical exponents and cross-check these measurements with constraints 
obtained from the theory of second order phase transitions.
For example, determining the exponent of the heat capacity $\alpha$ 
seems to be straight-forward, as the heat capacity is measured by
the event-by-event fluctuations of the chemical freeze-out temperature. 

\section{Conclusions}
The results summarized here are relevant for the low energy program
of the RHIC accelerator, to second or third generation experiments in the
heavy ion programme at  CERN SPS as well as for the cold and dense hadron matter physics 
planned at the forthcoming  FAIR facility. The methods outlined
can provide means to locate, if it exists, the critical end point
of the line of first order phase transitions in QCD. 
Based on an analogy with experiments in physical chemistry and
statistical physics, a novel method to determine opacity was proposed
and critical opalescence was suggested as a smoking gun experimental
signature for the vicinity of the critical point.
By measuring the critical exponents of the correlation function and the correlation
length with the outlined methods, 
the universality class of the QCD critical point can be determined.


\acknowledgments
It is my pleasure to thank P\'eter L\'evai for creating  
an inspiring atmosphere in Tokaj, and for his kindness during the finalization 
of this manuscript, and to M. Nagy and R. V\'ertesi for their careful reading of the manuscript.


\end{document}